\documentclass[aps,prb,floatfix]{revtex4}
\usepackage{times}
\usepackage{amsmath}
\usepackage{amsfonts}
\usepackage[]{graphicx}
\newcommand{\Sh}{Schr\"odinger{ }}

\newcommand{\bs}[1]{\boldsymbol{#1}}

\newcommand{\fues}[1]{\left(#1\right)}

\newcommand{\abs}[1]{\left\vert#1\right\vert}

\newcommand{\Eq}[1]{Eq. (\ref{#1})}

\begin{document}




\title[Microwave induced negative resistance states in 2D electron gas with periodic modulation]{Microwave induced negative resistance states in 2D electron gas with periodic modulation}


\author{Manuel Torres\footnote{Corresponding
     author: e-mail: {\sf torres@fisica.unam.mx.}}}
\affiliation{Instituto de F\'{\i}sica, Universidad Nacional
         Aut\'onoma de M\'exico, Apdo. Postal 20-364, M\'exico D.F.
	 01000, M\'exico}

\author{Alejandro  Kunold}
\affiliation{Departamento de Ciencias B\'asicas
         Universidad Aut\'onoma Metropolitana Azcapotzalco,
	 Av. San Pablo 180, Col. Reynosa Tamaulipas,
	 Azcapotzalco, M\'exico D.F. 02200, M\'exico}

\begin{abstract}
    We study the  microwave-induced   photoconductivity   of   a two-dimensional electron system (2DES)
  in the presence   of a magnetic field and a  two-dimensional modulation. The microwave and Landau contributions are exactly taken into account, while  the periodic potential is  treated perturbatively. The longitudinal resistivity exhibits oscillations, periodic in $\omega / \omega_c$.
Negative resistance  states  (NRS) develop  for sufficiently high  electron mobility and  microwave power.   This  phenomenon appears in a narrow window region of values of the lattice parameter ($a$),
around $a \sim l_B$, where $l_B$ is the magnetic length. 
It is proposed that these phenomena may be observed  in   artificially   fabricated arrays of periodic scatterers at the interface of  ultraclean  heterostructures.  

{73.20.At,05.60.-k, 72.15.Rn}
\end{abstract}
\maketitle                   




 Novel  strong  magnetoresistance  oscillations with the appearence  of   zero resistance states  (ZRS) were recently discovered
  \cite{mani1,zudov1},  when  ultraclean  $GaAs/Al_xGa_{1-x} As$ samples were   subjected to microwave irradiation and moderate magnetic fields.      
  It  is believed that the ZRS are probably originated from the evolution of negative resistance states (NRS) \cite{andre}.  
  Nowadays two distinct mechanisms that produce negative longitudinal resistance are known: 
  $(i)$  the impurity scattering mechanism \cite{theory1,tor0} and $(ii)$ the 
  inversion population   mechanism  \cite{theory2}. 
 Although the experiments described above do not include 
 the effect of   periodical potential modulations,   exploring its  physical consequences is worthwhile, see the references \cite{dietel}, 
  \cite{gumbs}  and  \cite{tor}.  In this work we make a theoretical study of the  microwave photoconductivity  of a  2DES in the presence of a magnetic field and a  two-dimensional modulation.

Consider the motion of an  electron in two dimensions subject to a uniform magnetic  field 
 $\mathbf{ B}$  perpendicular to the plane, a periodic potential $V$   and driven by  microwave radiation. The 
 dynamics is governed by the \Sh equation 
\begin{equation}\label{ecs1}
i \hbar \frac{\partial \Psi }{\partial t}= H \Psi  =  \left[  H_{\{B,\omega\}}  + V({\bs r} )  \right] \Psi  \, , \hskip1.5cm H_{\{B,\omega\}} = \frac{1}{2m^*} {\bs \Pi}^2 \, , 
\end{equation}
here $H_{\{B,\omega\}}$ is the Landau hamiltonian coupled to the radiation via the covariant derivative:  ${\bs \Pi} =  \mathbf{p} + e  \mathbf{A}$, with 
 $ \mathbf{A} = - \frac{1}{2} \bs r \times \bs B  +  Re \,\, \left[  \frac{\bs E_\omega }{\omega} \exp\{ -i \omega t \} \right]$.  
The  superlattice  potential   is  decomposed in a Fourier expansion:
$ V(\mathbf{r})= \sum_{m \, n}  V_{m \, n} \exp \bigg{\{i}  2 \pi  \left( \frac{m \, x}{a} + \frac{n \,  y}{b}   \right) \bigg { \}}. $
We shall assume:  $(i)$ a weak modulation    $ \vert  V \vert   \ll \hbar \omega_c $ and   $(ii)$
     the clean limit  $\omega \, \tau_{tr}   \sim  \omega_c \, \tau_{tr}  >>  1$; here $\tau_{tr}$  is the transport  relaxation time that is estimated using its relation to the   electron mobility $\mu = e \tau_{tr} / m^*$.  Based on these  conditions 
     it is justified to  consider the exact solution of the microwave driven Landau problem and treat   the periodic potential  effects  perturbatively.

A three step procedure is enforced  in order to solve the problem posed by \Eq{ecs1}: 
$(i)$  $H_{\{B,\omega\}}$   can be exactly diagonalized by a transformation of the form 
$W^{\dagger} H_{\{B,\omega\}}  W = \omega_c \left( \frac{1}{2} + a_1^\dagger \, a_1\right) \equiv H_0 $,
with the  $W(t)$ operator     given by   \cite{tor0} 
\begin{equation}\label{opw}
  W(t)=  \exp\{i \eta_1  \Pi_y\}  \exp\{i \zeta_1 \Pi_x\} \exp\{i \eta_2 {\mathcal O}_x\}  \exp\{i \zeta_2 {\mathcal O}_y\} \}   \exp\{i  \int^t  {\mathcal L} dt^\prime \} , 
 \end{equation}
where  the functions $\eta_i(t)$ and $\zeta_i(t)$ represent the solutions to the classical equations of motion associated with 
the Lagrangian $ {\mathcal L}  =  \frac{\omega_c}{2} \left( \eta_1^2 + \zeta_1^2 \right) +    \dot \zeta_1\eta_1 +  \dot \zeta_2 \eta_2 
+  e l_B \,    \left[ E_x  \left( \zeta_1 + \eta_2 \right)   +    E_y   \left(\eta_1 + \zeta_2  \right) \right]  $. The center-guide  operators 
are defined according to: $ \sqrt{e B} \,   {\mathcal O}_x   = \Pi_x  +  e By  \, , \hskip0.4cm \sqrt{e B}  \,  {\mathcal O}_y = \Pi_y  - e B x  $. 
These operators generate the  magnetic-translation algebra:  
$ \left[\Pi_i, \Pi_j \right]=i B \, \delta_{ij},   \hskip0.4cm  \left[  {\mathcal O}_i , {\mathcal O}_j \right]=- i B \, \delta_{ij}     \hskip0.4cm   
 \left[\Pi_i, {\mathcal O}_j \right]=0  $. 
$(ii)$  Applying the $W$-transformation,  the  \Sh \Eq{ecs1}  becomes 
$ i \hbar \frac{\partial \Psi^{(W)}  }{\partial t}  =  \left( H_0  + V_W (t)  \right) \Psi^{(W)},$ 
where  $ V_W  (t)  = W(t) V( {\bs r  }) W^{-1}(t)$ and $\Psi^{(W)}= W(t)  \Psi$. Notice that the periodic potential acquires a time dependence  brought by   the  $W(t)$ transformation. $(iii)$ The problem is now solved in terms of an evolution operator $U(t)$, using the  interaction representation  and first order time dependence perturbation theory: 
$U(t)  = 1 - i \int_{-\infty}^{t}  dt^\prime \left[ W^{\dagger}(t^\prime) V (\mathbf{r}) W(t^\prime)   \right]_I  \, .$  The solution to the original \Sh equation  has been achieved  by means of three successive  transformations
$ \vert  \Psi_\mu (t) \rangle  = W^\dag  \,  \exp\{-i H_0 t\} \,  U(t - t_0) \,  \vert  \mu \rangle $. The explicit expressions for the  matrix element of these  operators in the Landau-Floquet base  appear in detail 
in references \cite{tor0,tor}. 

\begin{figure}[htb]
\includegraphics[width=14cm, height=6.0cm]{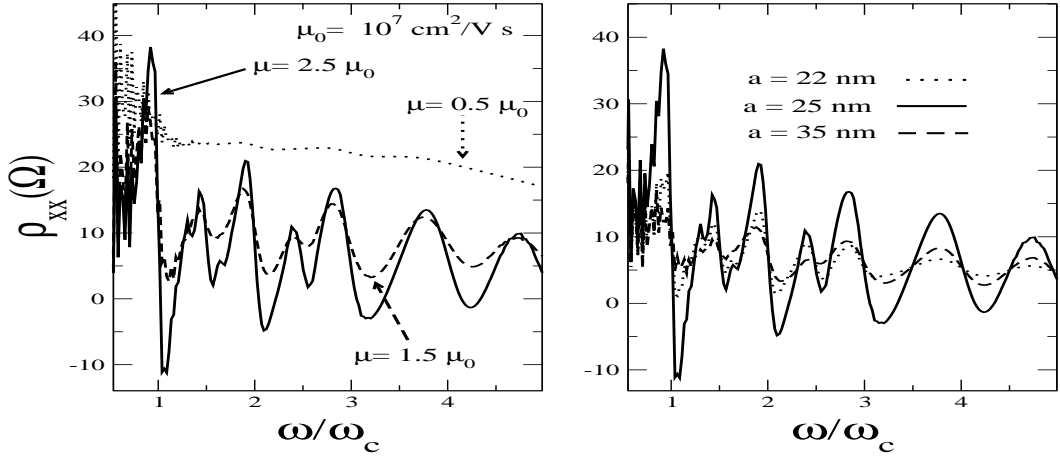}
\caption{(a)  Longitudinal resistivity  as a function $\omega/\omega_c$ for three  values
of the electron mobility: $\mu = 0.5 \times 10^7 \, cm^2/Vs  $   dotted line, $\mu = 1.5 \times 10^7 \, cm^2/Vs  $ dashed line, and 
 $\mu = 2.5 \times 10^7 \, cm^2/Vs  $  continuos line. In this case $a = 25 \, nm$. (b)  $\rho_{xx} \,$   versus $\omega/\omega_c$ 
for three  values of the lattice parameter:   $a=22 \, nm$ dotted line,  $a=25 \, nm$  continuos line,  and  $a=35 \, nm$    dashed  line; in this case $ \mu = 2.5 \times 10^7 cm^2/V  s$ .
  The microwave  polarization  is linear transverse (with respect to the current),  with  $f  = 100  \, Ghz$   and 
$\vert E_\omega \vert = 15 \, V/cm$.  The other parameters are selected as follows:
 $V_0=0.05 \, meV $,
$m^* = 0.067 \, m_e$,    
$\epsilon_F = 10 \, meV$, $T =1  \,  K$.   }
\label{figure1}
\end{figure}

The usual Kubo formula for the conductivity must be modified in order to include the Floquet dynamics  \cite{tor0}. 
 The longitudinal and Hall conductivities are separated in a dark and
   microwave induced contributions  \cite{tor}.   Here we  quote the result for the
  microwave induced  longitudinal photoconductivity 
 \begin{equation}\label{condLw}
\hskip-0.8cm  \mathbf{\sigma}_{xx} ^{(MM)}   =   \frac{ e^2  l_B^2 }{\pi \hbar}  
\int d {\cal E}  \sum_{\mu \nu}  \sum_l  \,   \sum_{m \, n}  \, Im G_\mu \left({\cal E}   \right)B^{(l)}   \left({\cal E}  ,{\cal E} _\nu \right)       \,
  \bigg{ \vert} q_n^{(y)} \,  J_l\left( \vert \Delta_{m \, n}  \vert \right)  V_{m \, n}   D_{\mu\nu} (\tilde{q}_{mn}) \bigg{\vert}^2  ,
 \end{equation}
where  $l_B=  \sqrt{\frac    {\hbar}{eB}}$ is the magnetic length, 
$  q^{(x)}_m = 2 \pi m /a $,  $ q^{(y)}_n= 2 \pi n /b  $, and $   B^{(l)}$ is given  by
 \begin{equation} \label{derspec} 
  B^{(l)} ( {\cal E} ,  {\cal E} _\nu) =  -  \frac{d}{d{\cal E} _0}  \bigg{ \{ } \left[ f( {\cal E} + l \omega + {\cal E} _0)-  f ( {\cal E} )\right]  
 Im \, G_\nu ( {\cal E}  + l \omega   + {\cal E} _0) \bigg{ \}}
  \bigg{\vert}_{{\cal E} _0= 0}. 
 \end{equation}
Broadening effects are included phenomenologically choosing a Gaussian representation for the density of states 
  \begin{equation} \label{denst}
  Im \, G_\mu({\cal E} ) = \sqrt{\frac{\pi}{2 \Gamma_\mu^2}}  \exp{\left[ - ({\cal E}  - {\cal E} _\mu)^2 /(2 \Gamma_\mu^2) \right] } ,  \hskip1.5cm    \Gamma^2_\mu = \frac{ 2\beta  \hbar^2 \omega_c  }{ (\pi \tau_{tr})} \, .  
 \end{equation}   
  In \Eq{condLw},   $J_l\left( \vert \Delta \vert \right)$ is the Bessel function with
$ \Delta =  \frac{\omega_c l_B^2 e E }{\omega \left( \omega^2 - \omega_c^2 + i \omega \Gamma \right)}
\left[ \omega \left(q_x e_x + q_y e_y  \right)  + i \omega_c \left(q_x e_y  -  q_y e_x  \right) \right]$, where   $\bs e $ is  the microwave polarization vector.
     $D^{\nu \mu}\fues{ \tilde{q}}= e^{-\frac{1}{2}\abs{  \tilde{q}}^2}  \tilde{q}^{\nu-\mu}\sqrt{\frac{\mu!}{\nu!}} L^{\nu-\mu}_{\mu}\fues{\abs{ \tilde{q}}^2},$ with  $L^{\mu}_{\mu}$ being the generalized Laguerre polynomial and  $\tilde{q} = i l_B (q_x - i q_y)/ \sqrt{2}$. 
The parameter  $\beta  \approx 10.5 $   takes into account the difference  of the transport scattering time $\tau_{tr}$   from the single-particle  lifetime $\tau_s$.  

In our calculations it is assumed  that a  superlattice  is cleaved   at the interface of an  ultraclean  \\ $GaAs/Al_xGa_{1-x} As$ heterostructure   with high  electron  mobility.  We shall consider a square lattice potential of the form 
   $  V(\mathbf{r}) = V_0 \left[ \cos\left( \frac{2 \pi x }{a} \right)  +  \cos\left( \frac{2 \pi y }{a} \right) \right] $. 
 Negative magnetoresistance requires ultra-clean  samples,  the phenomenon  appears when the 
   electron mobility exceeds  a threshold  $ \mu_{th}$.  Figure (1a)  displays $\rho_{xx} \, \, v.s. \,\, \omega / \omega_c $ plots for three  selected values of $\mu$. For $\mu \approx 0.5  \times 10^7 cm^2/V  s$  an   almost linear behavior $\rho_{xx} $ is clearly depicted (except in the Shubnikov-deHass region).  As the electron mobility  increases to $\mu \approx 1.5  \times 10^7 cm^2/V  s$, resistance  oscillations periodic in $\omega/ \omega_c$  are clearly observed. However,  several NRS    appear only when   the mobility is increased to $\mu \approx 2.5  \times 10^7 cm^2/V  s$. The explanation for the NRS can be traced down to 
 Eqs.    (\ref{condLw}) and  (\ref{derspec});  the  longitudinal photoconductivity contains a  new contribution proportional to the derivative of the density of states:  $   \frac{d}{d{\cal E} }  Im \, G_\nu ( {\cal E}  + l \omega )$.   Due to the oscillatory structure of the density of states this extra contribution takes both  positive and negative values. This  term is proportional to the electron mobility, hence  for sufficiently  high mobility the new contribution dominates over the dark contribution,  leading to  negative conductivity  states. 
 Figure (1b) shows that   NRS appear only in a narrow 
window of values of $a$ around   $a^*$, for which  the oscillations amplitude of $\rho_{xx}$  attains  its maximum. It is found 
that:  $a^* \sim \pi l_B / \sqrt{2} $.   Unlike the  semiclassical origin of magnetoresistance oscillations observed in  an antidot array for commensurate values of the ratio $R_c/a$,   these  conductance oscillations have a quantum origin and would  only appear in a narrow window of values  around
  $ a \sim l_B$. 
      
In conclusion, it is  proposed that  the   combined effects of: periodic modulation,  perpendicular magnetic field,  plus microwave  irradiation  of  2DES predicts the existence of  interesting oscillatory conductance phenomena, with the possible   development of  NRS.

 We acknowledge  financial support  by
CONACyT  and UNAM project No.  \texttt{IN113305}.

\end{document}